\documentclass[a4paper,11pt]{article}
\usepackage{color,xcolor,ucs}
\usepackage[top=0.3in, bottom=0.5in, left = 0.65in, right = 0.65in]{geometry}
\usepackage[linkcolor=black,colorlinks=true,urlcolor=blue]{hyperref}
\usepackage{mathtools}

\usepackage{color,xcolor,ucs}
\usepackage{mathtools}   \usepackage{tikz} 
\usepackage{ amssymb }
\usepackage{extarrows} 
\usepackage{pgf,tikz}
\usepackage{float}
\usetikzlibrary{positioning}
\usetikzlibrary{shapes.geometric}
\usetikzlibrary{shapes.misc}
\usetikzlibrary{arrows}
\usepackage{caption}
\usepackage{mathrsfs}
\usetikzlibrary{arrows,shapes,automata,backgrounds,petri,positioning}
\usetikzlibrary{decorations.pathmorphing}
\usetikzlibrary{decorations.shapes}
\usetikzlibrary{decorations.text}
\usetikzlibrary{decorations.fractals}
\usetikzlibrary{decorations.footprints}
\usetikzlibrary{shadows}
\usetikzlibrary{calc}
\usetikzlibrary{spy}
\usepackage{amsmath}
\usepackage{array}
\usepackage{ amssymb }
\usepackage{braket}
\usepackage{qcircuit}
\usepackage{soul}
\usepackage{braket} 
\usepackage{relsize}

\usepackage{amsmath}
\usepackage{ amssymb }
\usepackage{braket}
\usepackage{qcircuit}
\usepackage{soul}
\usepackage{braket}

\title{{\LARGE Eigenvalue attraction in open quantum systems, biophysical systems, and Parity-Time symmetric materials}}
\author{P. Rigas}
\date{}

\begin{document}

\maketitle

\begin{abstract}
  We investigate eigenvalue attraction for open quantum systems, biophysical systems, and for Parity-Time symmetric materials. To determine whether an eigenvalue and its complex conjugate of a real matrix attract, we derive expressions for the second derivative of eigenvalues, which is dependent upon contributions from inertial forces, attraction between an eigenvalue and its complex conjugate, as well as the force of the remaining eigenvalues in the spectrum. \footnote{\textit{{Keywords}}: 
  Real matrices, PT symmetric materials, biophysical systems, time evolution} \footnote{\textbf{MSC Class}: 81S02; 82D02; 81V02}
\end{abstract}

\section{Introduction}

\subsection{Overview}

\noindent Random matrices have emerged as an intense field of study in probability theory, with efforts devoted towards quantifying spectra {\color{blue}[6]}, distribution of eigenvalues {\color{blue}[3]}, information theory {\color{blue}[1]}, black holes {\color{blue}[2]}, the circular law {\color{blue}[7]}, and expected norms of random matrices {\color{blue}[10]}. To further explore some directions of interest relating to random matrices that are raised in {\color{blue}[4]}, we make use of a previously developed framework, from {\color{blue}[4]}, which provides three possible behaviors for attraction between eigenvalues of a real in the complex plane. First, we provide several expressions for the force of the complex conjugate of an eigenvalue on an eigenvalue, the expectation of this force, which has contributions from an inertial component, as well as contributions from interactions between eigenvalues of the spectrum. In the context of open quantum systems, {\color{blue}[4]}, notions of eigenvalue attraction arise from descriptions of a Hamiltonian and the Hilbert space in which time evolution occurs, while in the context of biophysical systems, {\color{blue}[5]}, notions of eigenvalue attraction arise in descriptions of ensemble behavior, which have connections with properties of spectra of eigenvalue distributions. Separate from open quantum and biophysical systems, Parity-time symmetric materials exhibit eigenvalue attraction from expressions that were previously obtained for eigenvalues of scattering matrices. Regardless of the specific environment in which eigenvalues, and spectra, of matrices are studied, second order differential equations for the evolution of eigenvalues within the spectra can be obtained, which not only raises implications for the forces that any two eigenvalues within the spectra exert on each other, but also for collisions between eigenvalues. To demonstrate how such formulas can be obtained for each system of interest, we distinguish between three possible forces that the eigenvalue of the spectra can exhibit, followed by an overview of expressions for the second order differential equation of the eigenvalues for open quantum systems, Biophysical systems, and Parity-time symmetric materials.

\subsection{Matrix objects}

\noindent For an $n \times n$ real matrix $M\big( t \big) \in \textbf{R}^{n \times n}$ parametrized at time $t$, the force $F$ between an eigenvalue and its complex conjugate takes the form,

\begin{align*}
\underset{j \in \textbf{N}}{\sum}  F \big( \bar{\lambda_j} \longrightarrow \lambda_j \big) =  - i  \underset{j \in \textbf{N}}{\sum}   \frac{\big|        u^{\mathrm{T}}_j \dot{M} \big( t \big) v_j     \big|}{\mathrm{Im} \big( \lambda_j \big( t \big) \big) }         \text{ } \text{ , } 
\end{align*}

\noindent for the standard right, and left, eigenvectors $v_i$ and $u_i$ of $\lambda_i$. Under the representation of $M \big( t \big)$ as a time-dependent stochastic process, $M \big( t_i + \delta t \big) = M \big( t_i \big) + \delta t P \big( t_i \big)$, for $\delta t \in \big[0,t_{i+1} - t_i \big]$, where $P \big( t_i \big)$ is a diagonal matrix at $t_i$. From the force between an eigenvalue and its complex conjugate, the expected value,

\begin{align*}
         \textbf{E} \big[   \underset{j \in \textbf{N}}{\sum}  F \big( \bar{\lambda_j} \longrightarrow \lambda_j \big)\big] \equiv  \underset{j \in \textbf{N}}{\sum}   \textbf{E} \big[  F \big( \bar{\lambda_j} \longrightarrow \lambda_j \big)  \big] \equiv  - i \underset{m,l \in \textbf{N} : m,l \neq j}{\underset{j \in \textbf{N}}{\sum} }             \frac{  \textbf{E} \big[ p^2_{ml} \big]  \big|   u^{*,m}_i   \big|^2 \big| v^l_i \big|^2   }{2 \text{ } \mathrm{Im} \big(   \lambda_j \big( t \big)     \big) }        \\ \overset{(\mathrm{iid})}{=} - i  \underset{j \in \textbf{N}}{\sum}  \frac{\textbf{E} \big[           p^2 \big] \big|\big|        u_j \big|\big|^2_2}{2 \text{ } \mathrm{Im} \big( \lambda_j \big( t \big) \big) }               \text{ } \text{ , } 
\end{align*}

\noindent with respect to the probability measure $\textbf{P}\big( \cdot \big)$ of standard normal random variables. For $\epsilon$ sufficiently small, as the parameter approaches $0$,

\begin{align*}
 M \big( t \big)  \equiv \underset{\epsilon \longrightarrow 0}{\mathrm{lim}}  M_{\epsilon} \big( t \big)  \text{ } \text{ . } 
\end{align*}

\noindent Following the discussion of {\color{blue}[6]}, the eigenvalue equations are,

\begin{align*}
M \big( t \big) v_i \big( t \big) = \lambda_i \big( t \big) v_i \big( t \big) \text{ } \text{ , } \\ u^{\mathrm{*}}_i \big( t \big) M \big( t \big) = \lambda_i \big( t \big) u^{\mathrm{*}}_i \big( t \big)         \text{ } \text{ , }
\end{align*}

\noindent in which, as in the previous objects, the left and right eigenvectors, and eigenvalues, of the time-dependent, real matrix $M \big( t \big)$ appear. From the eigenvalue equations, the matrix of eigenvectors,

\begin{align*}
V \big( t \big) \equiv   \underset{1 \leq i \leq n}{\bigcup}   {\mathrm{span}} \big\{ v_i \big( t \big)  \big\}    \text{ } \text{ , } 
\end{align*}

\noindent the Kronecker delta is obtained from the product,

\begin{align*}
 u^{\mathrm{*}}_j \big( t \big) v_i \big( t \big)  = \delta_{ij}  \text{ } \text{ . } 
\end{align*}

\noindent Differentiating the equations above involving $M \big( t \big)$, and the left and right eigenvectors, yields,

\begin{align*}
          \dot{M \big( t \big)} v_i \big( t \big) + M \big( t \big) \dot{v_i \big( t \big)} = \dot{\lambda_j \big( t \big) } v_i \big( t \big)     \text{ } \text{ , } \\    u^{\mathrm{*}}_i \big( t \big) + \dot{M \big( t \big)} + \dot{u^{\mathrm{*}}_i \big( t \big) } M \big( t \big) = \dot{\lambda_i \big( t \big) } u^{\mathrm{*}}_j \big( t \big) + \lambda_i \big( t \big)      \dot{u^{\mathrm{*}}_i \big( t \big)}     \text{ } \text{ , } 
\end{align*}  

\noindent which implies that the velocity, and second derivative with respect to time, of the eigenvalues are,

\begin{align*}
   \dot{\lambda_j \big( t \big)} = u^{\mathrm{*}}_j \big( t \big) \dot{M \big( t \big)} v_j \big( t \big)  \text{ } \text{ , } 
\end{align*}

\noindent and,

\begin{align*}
 \ddot{\lambda_j \big( t } \big) = u^{\mathrm{*}}_j \big( t \big)  \ddot{M \big( t \big)} + 2 \underset{i \neq j}{\sum} \frac{\big(             u^{\mathrm{*}}_i \big( t \big) \dot{M\big( t \big)}    v_j \big( t \big)  \big) \big(   u^{\mathrm{*}}_j \big( t \big) \dot{M \big( t\big)} v_i \big( t \big)       \big)   }{\lambda_i \big( t \big) - \lambda_j \big( t \big)} \text{ } \text{ , } 
\end{align*}

\noindent respectively, for matrices that are not Hermitian. In the case of circulant matrices, given diagonal $P \equiv \mathrm{diag} \big( p_1 , \cdots , p_n \big)$, the second order relation above is equivalent to,

\begin{align*}
  \ddot{\lambda_j \big( t } \big) =  - \frac{i}{\mathrm{Im} \big( \lambda_j \big( t \big) \big) } \big|   v^a_k p_a v^a_k           \big|^2 + 2 \underset{i\neq j}{\sum}   \frac{\big|       \bar{v}^a_k p_a v^a_j         \big|^2}{\lambda_i \big( t \big)  - \lambda_j \big( t \big)  }      \text{ } \text{ . } \tag{1}
\end{align*}

\noindent More generally, the expression above can also be expressed as,

\begin{align*}
  \ddot{\lambda_j \big( t } \big) =  u^{\mathrm{*}}_i \big( t \big) \ddot{M \big( t \big)} v_i \big( t \big) + 2 \underset{i \neq j}{\sum} c_{ij} c_{ji} \frac{\hat{r_{ij}}}{\big| r_{ij}\big|}  \text{ } \text{ , } 
\end{align*}

\noindent for,

\begin{align*}
 \hat{r_{ij}} \equiv  \frac{\bar{\lambda_i \big( t \big)} - \bar{\lambda_j \big( t\big)}}{\big| \lambda_i \big( t \big) - \lambda_j \big( t \big) \big| } \text{ } \text{ , } 
 \end{align*}

 \noindent and,

 \begin{align*}\big| r_{ij}\big| \equiv         \sqrt{\big| x_{ij} \big|^2 + \big| y_{ij} \big|^2 }      =   \sqrt{\big( \mathrm{Re} \big( \bar{\lambda_i \big( t \big)} + \mathrm{Re} \big( - \bar{\lambda_j \big( t \big)} \big)^2 + \big( \mathrm{Im} \big( \bar{\lambda_i \big( t \big)} + \mathrm{Im} \big( - \bar{\lambda_j \big( t \big)} \big)^2}         \text{ } \text{ . } 
\end{align*}

\subsection{Paper organization}

\noindent To further expand upon the eigenvalue attraction framework presented in the previous section, in the next section we present an overview of characteristics of eigenvalues of real matrices arising in descriptions of biophysical systems {\color{blue}[5]}.

\bigskip

\noindent In each setting, we provide closed expressions for $\ddot{\lambda_j \big( t \big)}$, which are each provided in the following \textit{Main Result}:

\bigskip

\noindent \textit{Main Result}, $\ddot{\lambda_j \big(t \big)}$ for open quantum systems, Biophysical systems, and Parity-time symmetric materials.

\begin{itemize}
    \item[$\bullet$] \textit{Case one}, \textit{second derivative of the eigenvalues for state matrices of open quantum systems}. For open quantum systems, $\ddot{\lambda_j \big( t \big)}$ reads,

\begin{align*}
  \bra{\bar{\Psi_j}} \ddot{\widetilde{H} \big( t \big) }                 + \underset{i\neq j}{\sum}       \frac{\big( \bra{\bar{\Psi_i}} \dot{\widetilde{H}  \big( t \big)} \ket{\Psi_j}  \big) \big(     \bra{\bar{\Psi_j}}   \dot{\widetilde{H}  \big( t \big)}  \ket{\Psi_i}   \big) }{h_i - h_j }             \text{ } \text{ . } 
\end{align*}

      \item[$\bullet$] \textit{Case two}, \textit{second derivative of the eigenvalues for state matrices of biophysical systems}. For biophysical systems, $\ddot{\lambda_j \big( t \big)}$ reads,

\begin{align*}
   \mathrm{exp} \big( - \frac{\big| \bar{\vec{r}} - \bar{\vec{r_j}} \big|}{\xi_n} \big) \ddot{\Omega^{\mathrm{LE}}}  \big( t \big) +   \underset{i \neq j}{\sum} \frac{      \big(   \mathrm{exp} \big( -  \frac{\big| \bar{\vec{r}} - \bar{\vec{r}_i} \big|}{\xi_n } \big)   \dot{\Omega^{\mathrm{LE}}}   \big( t \big)  \text{ } \mathrm{exp} \big( - \frac{\big| \vec{r} -  \vec{r}_j \big|}{\xi_n} \big) \big) \big(   \mathrm{exp} \big( -     \frac{\big| \bar{\vec{r}}- \bar{\vec{r}}_j  \big|}{\xi_n }                \big)      \dot{\Omega^{\mathrm{LE}}}  \big( t \big) \text{ } \mathrm{exp} \big( -    \frac{\big| \vec{r} -  \vec{r}_i \big|}{\xi_n }        \big)     \big) }{\Lambda_i - \Lambda_j }   \text{ . } 
\end{align*}

        \item[$\bullet$] \textit{Case three}, \textit{second derivative of the eigenvalues for state matrices of parity-time symmetric materials}. For parity-time symmetric materials, $\ddot{\lambda_j \big( t \big)}$ reads,

\begin{align*}
                \frac{\dot{\mathcal{P}} \big( j , i \big)}{v_j \big( t \big) }       +  \underset{i \neq j}{\sum}        \frac{   \mathcal{P} \big( i , j \big)  \mathcal{P} \big( j , i \big) }{\frac{1\pm \sqrt{1- M_{11}( i  ) M_{22}(i)}}{M_{22} ( i ) } - \frac{1\pm \sqrt{1- M_{11}( j  ) M_{22}(j)}}{M_{22} ( j ) } 
 }              \text{ } \text{ . } 
\end{align*}

\end{itemize}

\noindent For each of the cases above, $\widetilde{H}$, $\Omega^{\mathrm{LE}}$, and $M$ denote the states matrices, $\dot{\widetilde{H}}$, $\dot{\Omega^{\mathrm{LE}}}$, and $\dot{M}$ denote the first time derivatives of the states matrices, $\ddot{\widetilde{H}}$, $\ddot{\Omega^{\mathrm{LE}}}$, and $\ddot{M}$ denote the second time derivatives of the states matrices for open quantum systems, biophysical systems, and parity-time symmetric materials, respectively. For the remaining quantities, $\bra{\widetilde{\Psi_j}}$, and $ \mathrm{exp} \big( - \frac{\big| \bar{\vec{r}} - \bar{\vec{r_j}} \big|}{\xi_n} \big)$ denote the $j$ th left eigenvectors, $\bra{\widetilde{\Psi_i}}$, and $ \mathrm{exp} \big( - \frac{\big| \bar{\vec{r}} - \bar{\vec{r_i}} \big|}{\xi_n} \big)$, denote the $i$ th left eigenvectors, $\ket{\Psi_j}$, and $\mathrm{exp} \big( - \frac{\big| \bar{\vec{r}} -  \bar{\vec{r}_j} \big|}{\xi_n} \big)$, denote the $j$ th right eigenvectors, and $\ket{\Psi_i}$, and $\mathrm{exp} \big( - \frac{\big| \bar{\vec{r}} -  \bar{\vec{r}_i} \big|}{\xi_n} \big)$ denote the $i$ th right eigenvectors, for open quantum systems, and for biophysical systems, respectively. For parity-time symmetric materials, $\mathcal{P} \big( i , j \big)$ and $\mathcal{P} \big( j , i \big)$ respectively denote $           u^{\mathrm{*}}_i \big( t \big) \dot{M\big( t \big)}    v_j \big( t \big)$, and $  u^{\mathrm{*}}_j \big( t \big) \dot{M \big( t\big)} v_i \big( t \big)$. Finally, from each of the three cases above, $h_i$, $\Lambda_i$, and $\frac{1\pm \sqrt{1- M_{11}( i  ) M_{22}(i)}}{M_{22} ( i ) } $ denote the $i$ th eigenvalue of open quantum systems, biophysical systems, and parity-time symmetric materials, while, similarly, $h_j$, $\Lambda_j$ and $\frac{1\pm \sqrt{1- M_{11}( j  ) M_{22}(j)}}{M_{22} ( j ) } $ denote the $j$ th eigenvalues.

\section{Eigenvalue attraction}

\noindent We apply the framework described in {\color{blue}[4]} to Open Quantum systems, Biophysical systems, and Parity-Time symmetric materials below.

\subsection{Open quantum systems}

\noindent The first application is to open quantum systems.

\subsubsection{Description}

\noindent In the open quantum system setting, {\color{blue}[9]}, from the time evolution of $M \big( t \big)$, the procedure for obtaining the eigenvalues, and eigenvectors, of a real matrix differs. Instead, for Hamiltonians that are not Hermitian, the eigenvalues and eigenvectors are similarly observed to abruptly change, as we have described for Biophysical systems. To determine locality constraints for pure state preparation without undesired decoherence and/or interference, quantum dynamical semigroups are generated by the Markovian Master Equation,

\begin{align*}
  \frac{\mathrm{d} \rho \big( t \big) }{\mathrm{d} t } =  - i \big[ H , \rho \big( t \big) \big] + \sum_k \big(   L_k \rho \big( t \big) L^{\dagger}_k - \frac{1}{2} \big\{ L^{\dagger}_k L_k , \rho \big( t \big) \big\}       \big)   \text{ } \text{ , } 
\end{align*}

\noindent for $H = H^{\dagger}$, where the time evolution of the density $\rho \big( t \big)$ at $t$ is equivalent to contributions from an imaginary term of the Lie bracket between $H$ and $\rho \big( t \big)$, as well as a summation over $k$ of the Lindblad operators $\big\{ L_k \big\}$. From this relationship between the time evolution and Hamiltonian, from a decomposition of the Hilbert space over $n$ components, $\mathcal{H}_{\mathscr{Q}} = \otimes_{a=1}^n \mathcal{H}_a$, the Hamiltonian is said to be \textit{QL} if,

\begin{align*}
  H = \sum_j H_j   \text{ } \text{ , } 
\end{align*}

\noindent with each $H_j = H_{\mathcal{N}_j} \otimes I_{\bar{\mathcal{N}_j}}$, for,

\begin{align*}
  I_{\bar{\mathcal{N}_j}} = \otimes_{a \not\in \mathcal{N}_j} I_a   \text{ } \text{ , } 
\end{align*}

\noindent and $\mathcal{N}_j \subseteq \big\{ 1 , \cdots , n \big\}$, for $j = 1 , \cdots , M$, and identity operators $I$ and $I_a$, under the partition,

\begin{align*}
     \mathcal{H}_{\mathscr{Q}} =  \mathcal{H}_{S} \oplus \mathcal{H}^{\perp}_{S}       \text{ } \text{ , } 
\end{align*}

\noindent for the block representation,

\[
X   \equiv 
  \begin{bmatrix}
   X_S   &  X_P \\   X_Q &   X_R   \text{ }  
  \end{bmatrix} \text{ } \text{ , } 
\]

\noindent From this notion of the Hamiltonian being \textit{QL}, the Markov Master Equation is said to be \textit{QL} if both the Hamiltonian and $\big\{ L_k \big\}$, the noise operators, are \textit{QL}. The Hamiltonian and noise operators are said to be local if each can be expressed with an identity operator with the exception of at most one subsystem.

\bigskip

\noindent Next, the related notion of how a pure quantum state can be efficiently prepared asymptotically, termed the globally asymptotically stable state, entails that for some initial configuration $\rho_0$, the limit in infinite time,

\begin{align*}
    \underset{t \longrightarrow + \infty}{\mathrm{lim}}   \mathrm{exp} \big( \mathcal{L} t \big) \big( \rho_0 \big) = \rho             \text{ } \text{ , } 
\end{align*}

\noindent stabilizes to $\rho$, independently of the initial condition, for,

\begin{align*}
   \mathrm{exp} \big( \mathcal{L} t \big) \equiv \overset{n}{\underset{a=1}{\otimes}} \mathrm{exp} \big( \mathcal{L}_a t \big) \text{ } \text{ . } 
\end{align*}

\noindent If the state that we wish to prepare in an asymptotically, global, manner is not stable, it is said to otherwise be dissipatively quasi-locally stabilizable, (\textbf{Definition} \textit{2.1}, {\color{blue}[9]}), if,

\begin{align*}
     \frac{\mathrm{d} \rho }{\mathrm{d} t} = \sum_k \big( D_k \rho  D^{\dagger}_k - \frac{1}{2} \big\{ D^{\dagger}_k D_k , \rho \big\} \big)  \text{ } \text{ , } 
\end{align*}

\noindent is satisfied for the collection of \textit{QL} operators $\big\{ D_k \big\}$.

\subsubsection{Eigenvalue attraction statement}

 \noindent \textit{Main result, case one}. To characterize eigenvalue attraction for random matrices of Open Quantum systems, consider eigenvalues of the form, {\color{blue}[9]}, 

 \begin{align*}
       h  \approx           \widetilde{H}_S    \text{ } \text{ , } 
 \end{align*}

 \noindent which is related to the following block representation for $H$, similar to that introduced in \textit{2.2.1}, from,
 
 \[
H  \equiv 
  \begin{bmatrix}
   H_S   &  H_P \\   H_Q &   H_R   \text{ }  
  \end{bmatrix} \text{ } \text{ , } 
\]

 \noindent which holds from the fact that the Markovian Master Equation,

\begin{align*}
  \frac{\mathrm{d} \rho \big( t \big) }{\mathrm{d} t } =  - i \big[ H , \rho \big( t \big) \big] + \sum_k \big(   L_k \rho \big( t \big) L^{\dagger}_k - \frac{1}{2} \big\{ L^{\dagger}_k L_k , \rho \big( t \big) \big\}       \big)   \text{ } \text{ , } 
\end{align*}

\noindent is invariant under $\widetilde{H}$ and $\widetilde{L_k}$, in which,

  \begin{align*}
      \frac{\mathrm{d} \rho \big( t \big) }{\mathrm{d} t } =  - i \big[ \widetilde{H} , \rho \big( t \big) \big] + \sum_k \big(   \widetilde{L_k} \rho \big( t \big) \widetilde{L^{\dagger}_k} - \frac{1}{2} \big\{ \widetilde{L^{\dagger}_k} \widetilde{L_k} , \rho \big( t \big) \big\}       \big)     \text{ } \text{ , } 
 \end{align*}

 \noindent also holds, for,

 \begin{align*}
        \widetilde{H}  \equiv H + \frac{i}{2} \sum_k \big( l^{*}_k L_k - l_k L^{\dagger}_k  \big)    \text{ } \text{ , } 
 \end{align*}

 \noindent for $\widetilde{L_k} = L_k - l_k I$, with block diagonal $\widetilde{H}$, in which, for an arbitrary number of blocks along the diagonal,

 \[
\widetilde{H_S}  \equiv 
  \begin{bmatrix}
   \text{Block}   &  0 \\ \ddots & \ddots  \\  0 &   \text{Block}   \text{ }  
  \end{bmatrix} \text{ } \text{ , } 
\]

 \noindent With the eigenvectors being $\ket{\Psi}$, this quantum state is used to construct pure states $\ket{\Psi} \bra{\Psi}$ with the operators $\big\{ D_k \big\}$.

 \bigskip

 \noindent Equipped with the eigenvalues and eigenvectors, $\ddot{\lambda_j \big( t } \big)$ reads,

 \begin{align*}
    \ddot{\lambda_j \big( t } \big)  =  \bra{\bar{\Psi_j}} \ddot{\widetilde{H} \big( t \big) }                 + \underset{i\neq j}{\sum}       \frac{\big( \bra{\bar{\Psi_i}} \dot{\widetilde{H}  \big( t \big)} \ket{\Psi_j}  \big) \big(     \bra{\bar{\Psi_j}}   \dot{\widetilde{H}  \big( t \big)}  \ket{\Psi_i}   \big) }{h_i - h_j }                  \text{ } \text{ , } 
\end{align*}

\noindent where $h_i$ and $h_j$ respectively denote the $i$ th, and $j$ th, eigenvalues of $\widetilde{H}$, $\dot{\widetilde{H}}$ denotes the first time derivative of the state matrix $\widetilde{H}$,

 \begin{align*}
        \dot{\widetilde{H}}  \equiv \dot{H} + \frac{i}{2} \sum_k \big( l^{*}_k \dot{L_k} - l_k \dot{L^{\dagger}_k}  \big)    \text{ } \text{ , } 
 \end{align*}

\noindent and the state $\bra{\widetilde{\Psi_i}}$ denotes the conjugate-transform of the state $\ket{\Psi_i}$.

\subsection{Biophysical systems}

\noindent The second applicaton is to biophysical systems.

\subsubsection{Description}

\noindent From the time evolution of $M$ with respect to $t$ that is defined in the previous section, in biophysical systems, as discussed in {\color{blue}[5]}, one encounters abrupt changes in the eigenvalues of real matrices, in which eigenvalues with positive real part can transition to have vanishing imaginary part. From standard models of diffusion underlying several biological processes, the linearized evolution matrix,

\[
\Omega  \equiv 
  \begin{bmatrix}
       a - 2D   &  D & \cdots & \cdots  & D  \\
    D  & a - 2 D & D & \cdots & 0 \\ 0 & D & \cdots & \cdots & 0 \\ \vdots & 0 & \vdots & \vdots & D \\   D & \cdots & 0 & D & a - 2 D       \text{ }  
  \end{bmatrix} \text{ } \text{ , } 
\]

\noindent arises from the approximation,

\begin{align*}
  \frac{\mathrm{d}c_n \big( t \big)  }{\mathrm{d} t }  \approx   \underset{m \in \textbf{N}}{ \sum}  \Omega_{nm} c_m \big( t \big)  \text{ } \text{ , } 
\end{align*}

\noindent where the function $c_n \big( t \big)$, the micro organism concentration per volume at time $t$, satisfies the discretization,

\begin{align*}
   \frac{\mathrm{d} c_n \big( t \big)}{\mathrm{d} t} = D \big( c_{n+1} + c_{n-1} - 2 c_n \big) + a c_n - b c^2_n \text{ } \text{ , } 
\end{align*}

\noindent obtained from the diffusion PDE,

\begin{align*}
  \frac{\partial c \big( x , t \big) }{\partial t}  = D \triangledown^2 c \big( x , t \big) + a c \big( x , t \big) - b c\big( x , t \big)^2 \text{ } \text{ , } 
\end{align*}

\noindent for the spatial diffusion constant $D$, which is taken to be strictly positive. Given some positive concentration of micro organisms per volume, over a strictly positive number $M$ of sites over the lattice, the linearized evolution matrix similarly reads,

\[
\Omega^{\mathrm{LE}}  \equiv 
  \begin{bmatrix}
       a - 2D  + U_1  &  D \mathrm{exp} \big( h \big) & \cdots & \cdots  & D \mathrm{exp} \big( - h \big)  \\
    D \mathrm{exp} \big( -  h \big)  & a - 2 D + U_2 & D \mathrm{exp} \big( h \big)  & \cdots & 0 \\ 0 & D \mathrm{exp} \big( - h \big)  & \cdots & \cdots & 0 \\ \vdots & 0 & \vdots & \vdots & D \mathrm{exp} \big( h \big)  \\   D \mathrm{exp} \big( h \big)  & \cdots & 0 & D \mathrm{exp} \big( - h \big) & a - 2 D  + U_N     \text{ }  
  \end{bmatrix} \text{ } \text{ , } 
\]

\noindent for fluctuations $\big\{ U_i \big\}_{1 \leq i \leq N}$ in the growth rate at site $i$, $b>0$, and the velocity flow field $\vec{v_0} \propto \vec{h}$. 

\subsubsection{Eigenvalue attraction statement}

\noindent \textit{Main result, case two}. To characterize eigenvalue attraction for random matrices of Biophysical systems, consider the collection of $n$ eigenvalues, {\color{blue}[5]}, $\Lambda_n$, with eigenfunctions,

\begin{align*}
 \psi \big( r , t \big) \equiv \sum_n c_n \psi_n \big( r \big) \mathrm{exp} \big( \Lambda_n t \big)    \text{ } \text{ , } 
\end{align*}

\noindent which are, given localization of each eigenfunction for $\vec{r}$ near $\vec{r}_n$, and inversely proportional to the localization length $\xi_n$,

\begin{align*}
   \psi_n \big(  \vec{r} , \vec{r}_n     \big) \overset{\vec{r} \approx \vec{r}_n}{\equiv} \psi_n \big( \vec{r} \big) \equiv \psi_n \big( \vec{r}_n \big)  \sim  \mathrm{exp} \big( - \frac{\big| \vec{r} - \vec{r}_n \big|}{\xi_n } \big) \text{ } \text{ , } 
\end{align*}

\noindent and, for a potential $U$. If one gathers data pertaining to the correlation length for each component, with $\big( \xi^1_n , \cdots , \xi^n_n \big)$, set $\xi_n \equiv \mathrm{sup}_i \xi^i_n$. Over all $n$, $\Lambda_n$ is equivalent to the union,

\begin{align*}
\Lambda \equiv \big\{ \text{set of eigenvalues of the operator } D  \triangledown^2    + U \big( r \big)   \big\} \equiv \underset{n}{\bigcup} \text{ } \Lambda_n \text{ } \text{ , } 
\end{align*}

\noindent where,

\begin{align*}
  \Lambda_n  \equiv \big\{ n \text{ th}  \text{\text{ } eigenvalue of the operator } D  \triangledown^2    + U \big( r \big)  \big\} \text{ } \text{ . } 
\end{align*}

\noindent Equipped with the eigenvalues and eigenvectors, $\ddot{\lambda_j \big( t } \big)$ reads,

\begin{align*}
      \big[   \mathrm{exp} \big( - \frac{\big|  \bar{r_1} - \bar{r^1_j} \big|}{\xi_n} \big) ,        \overset{n-2}{\cdots}  ,   \mathrm{exp} \big( - \frac{\big|  \bar{r_n} - \bar{r^n_j} \big|}{\xi_n} \big) \big]    \ddot{\Omega^{\mathrm{LE}}}   \big( t \big) \equiv  \mathrm{exp} \big( - \frac{\big| \vec{r} - \vec{r_j} \big|}{\xi_n} \big) \ddot{\Omega^{\mathrm{LE}}}  \big( t \big) \text{ } \text{ , } 
    \end{align*}

     \noindent for the first term, and, similarly,  
     
     \begin{align*}\underset{i\neq j}{\sum}  \frac{      \big[ \mathrm{exp} \big( - \frac{\big| \bar{r_1} - \bar{r^1_i}  \big|}{\xi_n }           \big) , \overset{n-2}{\cdots} ,  \mathrm{exp} \big( - \frac{\big| \bar{r_n} - \bar{r^n_i}  \big|}{\xi_n }  \big) \big] \dot{\Omega^{\mathrm{LE}} }   \big( t \big)    \big[ \mathrm{exp} \big( - \frac{\big| \bar{r_1} - \bar{r^1_j} \big|}{\xi_n}   \big) , \overset{n-2}{\cdots} ,  \mathrm{exp} \big( - \frac{\big| \bar{r_j} - \bar{r^n_j} \big|}{\xi_n} \big) \big]        }{\Lambda_i - \Lambda_j }        \times \cdots \\ \bigg( \big[    \mathrm{exp} \big( - \frac{\big| \bar{r_1} - \bar{r^1_j} \big|}{\xi_n } \big)   , \overset{n-2}{\cdots} ,       \mathrm{exp} \big( - \frac{\big| \bar{r_n}   - \bar{r^n_j} \big|}{\xi_n } \big)      \big] \dot{\Omega^{\mathrm{LE}}}  \big( t \big)  \big[     \mathrm{exp} \big( - \frac{\big| \bar{r_1} -  \bar{r^1_i} \big|}{\xi_n} \big) , \overset{n-2}{\cdots} ,    \mathrm{exp} \big( - \frac{\big| \bar{r_n} - \bar{r^n_i} \big|}{\xi_n} \big)      \big] \bigg)                  \text{ } \text{ , } 
\end{align*}

\noindent from the summation,

\begin{align*}
  \underset{i \neq j}{\sum} \frac{      \bigg(   \mathrm{exp} \big( -  \frac{\big| \bar{\vec{r}} - \bar{\vec{r}_i} \big|}{\xi_n } \big)   \dot{\Omega^{\mathrm{LE}}}   \big( t \big)  \text{ } \mathrm{exp} \big( - \frac{\big| \vec{r} -  \vec{r}_j \big|}{\xi_n} \big) \bigg) \bigg(   \mathrm{exp} \big( -     \frac{\big| \bar{\vec{r}}- \bar{\vec{r}}_j  \big|}{\xi_n }                \big)      \dot{\Omega^{\mathrm{LE}}}  \big( t \big) \text{ } \mathrm{exp} \big( -    \frac{\big| \vec{r} -  \vec{r}_i \big|}{\xi_n }        \big)     \bigg) }{\Lambda_i - \Lambda_j }  \text{ } \text{ , } 
\end{align*}

\noindent for the second term, where $\Lambda_i$ and $\Lambda_j$, respectively denote the $i$ th and $j$ th eigenvalues, $\dot{\Omega^{\mathrm{LE}}}$ is given by the matrix,

\[
\dot{\Omega^{\mathrm{LE}}}  \equiv 
  \begin{bmatrix}
      \dot{U_1}  &  D \dot{\mathrm{exp} \big( h \big)}  & \cdots & \cdots  & D \dot{\mathrm{exp} \big( - h \big) } \\
    D \dot{\mathrm{exp} \big( -  h \big)}  &  \dot{U_2} & D \dot{\mathrm{exp} \big( h \big)}  & \cdots & 0 \\ 0 & D \dot{\mathrm{exp} \big( - h \big)}  & \cdots & \cdots & 0 \\ \vdots & 0 & \vdots & \vdots & D \dot{\mathrm{exp} \big( h \big)}  \\   D \dot{\mathrm{exp} \big( h \big) } & \cdots & 0 & D \dot{\mathrm{exp} \big( - h \big)} &  \dot{U_N}     \text{ }  
  \end{bmatrix} \text{ } \text{ , } 
\]

\noindent $\ddot{\Omega^{\mathrm{LE}}}$ denotes the second time derivative of the $\Omega^{\mathrm{LE}}$, which is given by the matrix,

\begin{align*}
 \frac{\partial}{\partial t} \dot{\Omega^{\mathrm{LE}}}  \big( t \big) \equiv \ddot{\Omega^{\mathrm{LE}}}    \big( t \big) \equiv \ddot{\Omega^{\mathrm{LE}}}   \text{ } \text{ , } 
\end{align*}

\noindent and the $i$ th and $j$ th eigenvectors are constructed from the basis functions $\psi_n \big( \vec{r} , \vec{r}_n \big)$. Altogether, the second derivative of $\lambda_j \big( t \big)$ reads,

\begin{align*}
 \mathrm{exp} \big( - \frac{\big| \bar{\vec{r}} - \bar{\vec{r_j}} \big|}{\xi_n} \big) \ddot{\Omega^{\mathrm{LE}}}  \big( t \big) +   \underset{i \neq j}{\sum} \frac{      \bigg(   \mathrm{exp} \big( -  \frac{\big| \bar{\vec{r}} - \bar{\vec{r}_i} \big|}{\xi_n } \big)   \dot{\Omega^{\mathrm{LE}}}   \big( t \big)  \text{ } \mathrm{exp} \big( - \frac{\big| \vec{r} -  \vec{r}_j \big|}{\xi_n} \big) \bigg) \bigg(   \mathrm{exp} \big( -     \frac{\big| \bar{\vec{r}}- \bar{\vec{r}}_j  \big|}{\xi_n }                \big)      \dot{\Omega^{\mathrm{LE}}}  \big( t \big) \text{ } \mathrm{exp} \big( -    \frac{\big| \vec{r} -  \vec{r}_i \big|}{\xi_n }        \big)     \bigg) }{\Lambda_i - \Lambda_j }  \text{ } \text{ . } 
\end{align*}

\subsection{Parity-Time symmetric materials}

\noindent The third applicaton is to parity-time symmetric materials.

\subsubsection{Description}

\noindent In the PT symmetric materials setting, {\color{blue}[2,5]}, the eigenvalues of real matrices arise from considering an M-matrix and its connection with the scattering, S-matrix, in which,

\begin{align*}
        \vec{E}^{+} = M \vec{E^{-} }       \text{ } \text{ , } 
\end{align*}

\noindent for $\vec{E}^{+} = \big[ E^{+}_f E^{+}_B \big]^{\mathrm{T}}$, $\vec{E}^{-} = \big[       E^{-}_f E^{-}_B \big]^{\mathrm{T}}$, and,

\[
M   \equiv 
  \begin{bmatrix}
  M_{11}  &  M_{12} \\   M_{21} & M_{22}  \text{ }  
  \end{bmatrix} \text{ } \text{ , } 
\]

\noindent from properties of the transfer matrix $M$ above, {\color{blue}[2]}, which is given by the block representation,

\[
S   \equiv 
  \begin{bmatrix}
 T^l & R^r \\ R^l  &  T^r \text{ }  
  \end{bmatrix} \text{ } \text{ , } 
\]

\noindent where the entries of the scattering matrix above are given by, $T^l = \big( M_{22} \big( k \big) \big)^{-1}$, $R^r = \frac{M_{12} ( k )}{M_{22} ( k )}$, $R^l = - \frac{M_{21}(k)}{M_{22}(k)}$, and $T^r = \big( M_{22} \big( k \big) \big)^{-1}$.

\subsubsection{Eigenvalue attraction statement}

\noindent \noindent \textit{Main result, case three}. To characterize eigenvalue attraction for random matrices of PT symmetric materials, from unimodular $M$ with determinant $1$, consider the eigenvalues of the scattering matrix, which are given by,

\begin{align*}
     \lambda_{\pm,k}   \equiv s_{\pm,k} \equiv \frac{ 1 \pm  \sqrt{1 - M_{11} \big( k \big) M_{22} \big( k \big) }}{M_{22} \big( k \big) }             \text{ } \text{ . } 
\end{align*}

\noindent With each $\lambda_k$, the parity-time symmetric material satisfies the boundary conditions, in which

\begin{align*}
       \psi_{k_{\pm}} \big( x \big) = A_{\pm} \mathrm{exp} \big( i k_{\pm} x \big) + B_{\pm} \mathrm{exp} \big( - i k_{\pm} x \big) \overset{x \longrightarrow \pm \infty}{\longrightarrow }\mathrm{exp} \big(   \pm i k x        \big) 
       \text{ } \text{ , }
\end{align*}

\noindent are given by the linear combination of exponentials, with powers $i k_{\pm} x$, or $-i k_{\pm} x$. Asymptotically, for large $x$, the $k$ th left and right eigenvectors obey,

\[
\psi^{\mathrm{L}}_k \big( x \big) \equiv \psi^{\mathrm{L}}_k  \sim  \text{ } 
\left\{\!\begin{array}{ll@{}>{{}}l}  N_l \big( \mathrm{exp} \big( i k x \big) + R^l \mathrm{exp} \big( - i k x \big) \big)       &  \text{, as } x \longrightarrow - \infty 
 \text{ } \text{ , } \\   
        N_l T^l \mathrm{exp} \big( i k x \big)      &    \text{, as } x \longrightarrow +  \infty  \text{ }  \text{ , }  \\
\end{array}\right.
\]

\[
\psi^{\mathrm{R}}_k \big( x \big) \equiv \psi^{\mathrm{R}}_k  \sim  \text{ } 
\left\{\!\begin{array}{ll@{}>{{}}l}   N_r T^r \mathrm{exp} \big( - i k x \big)      &  \text{, as } x \longrightarrow - \infty 
 \text{ } \text{ , } \\   
         N_r \big(   \mathrm{exp} \big( - i k x \big) + R^r \mathrm{exp} \big( i k x \big) \big)    &    \text{, as } x \longrightarrow +  \infty  \text{ }  \text{ , }  \\
\end{array}\right.
\]

\bigskip

\noindent for complex valued coefficients $N_l$, $R^l$, $N_r$, $T^r$, and $R^r$. Equipped with the eigenvalues and eigenvectors, $\ddot{\lambda_j \big( t } \big)$ reads, from,

\begin{align*}
  \lambda_{\pm, i } -  \lambda_{\pm, j } = \frac{1\pm \sqrt{1- M_{11}( i  ) M_{22}(i)}}{M_{22} ( i ) } - \frac{1\pm \sqrt{1- M_{11}( j  ) M_{22}(j)}}{M_{22} ( j ) }  \text{ } \text{ , } 
\end{align*}

\noindent as,

\begin{align*}
    \ddot{\lambda_j \big( t } \big)   =                  \frac{\dot{\mathcal{P}} \big( j , i \big)}{v_j \big( t \big) }       +  \underset{i \neq j}{\sum}        \frac{   \mathcal{P} \big( i , j \big)  \mathcal{P} \big( j , i \big) }{\frac{1\pm \sqrt{1- M_{11}( i  ) M_{22}(i)}}{M_{22} ( i ) } - \frac{1\pm \sqrt{1- M_{11}( j  ) M_{22}(j)}}{M_{22} ( j ) } 
 }            \text{ } \text{ , } 
\end{align*}

\noindent where,

\begin{align*}
  \mathcal{P} \big( i , j , \dot{M } , u_i , v_j  \big)   \equiv   \mathcal{P} \big( i , j \big)  =        \big[  N_l \mathrm{exp} \big( i k {x_1} \big) + N_l R^l \mathrm{exp} \big( - i k {x_1}  \big)     , \overset{n-2}{\cdots}  ,   N_l  \mathrm{exp} \big( i k {x_n} \big) + \cdots \\ N_l  R^l \mathrm{exp} \big( - i k {x_n}  \big)                    \big] \dot{M \big( t \big) } \text{ } \times \cdots \\  \big[         N_r 
 \mathrm{exp} \big( i k {x_1} \big) + N_r  R^r \mathrm{exp} \big( - i k {x_1}  \big)        ,   \overset{n-2}{\cdots} ,           N_r \mathrm{exp} \big( i k_{\pm} {x_n} \big) + \cdots \\ N_r  R^r  \mathrm{exp} \big( - i k {x_n}  \big)   \big]                                 \text{ } \text{ , } 
\end{align*}

\noindent in the first term of the summation over $i\neq j$, for the $i$ th left eigenvector, 

\begin{align*}
\psi^{\mathrm{L},i}_k   \equiv   \big[  N_l  \mathrm{exp} \big( i k {x_1} \big) + N_l R^l \mathrm{exp} \big( - i k {x_1}  \big)     , \overset{n-2}{\cdots}  ,   N_l  \mathrm{exp} \big( i k {x_n} \big) +  N_l R^l  \mathrm{exp} \big( - i k {x_n}  \big)                    \big]    \text{ } \text{ , } 
\end{align*}

\noindent and for the $j$ th right eigenvector,

\begin{align*}
\psi^{\mathrm{R},i}_k   \equiv     \big[         N_r  \mathrm{exp} \big( i k {x_1} \big) + N_r R^r  \mathrm{exp} \big( - i k {x_1}  \big)        ,   \overset{n-2}{\cdots} ,           N_r \mathrm{exp} \big( i k {x_j} \big) +  N_r R^r  \mathrm{exp} \big( - i k {x_j}  \big)   \big]                            \text{ } \text{ . } 
\end{align*}

\noindent Similarly, for the other term in the summation over $i \neq j$,

\begin{align*}
    \mathcal{P} \big( j ,  i , \dot{M } , u_i , v_j  \big)   \equiv    \mathcal{P} \big( j , i  \big)  =         \big[  N_l \mathrm{exp} \big( i k {x_1} \big) + N_l R^l \mathrm{exp} \big( - i k {x_1}  \big)     , \overset{n-2}{\cdots}   ,      N_l \mathrm{exp} \big( i k {x_n} \big) + N_l R^l  \mathrm{exp} \big( - i k 
 {x_n}  \big)         \big] \times \cdots \\  \dot{M \big( t \big) } \big[  N_r  \mathrm{exp} \big( i k {x_1} \big) + N_r R^r  
 \mathrm{exp} \big( - i k {x_1}  \big)    , \overset{n-2}{\cdots} , \\           N_r  \mathrm{exp} \big( i k {x_n} \big) + N_r R^r \mathrm{exp} \big( - i k {x_n}  \big)       \big]                       \text{ } \text{ , } 
\end{align*}

\noindent in the second term of the summation over $i\neq j$, and,

\begin{align*}
 \frac{\dot{\mathcal{P}} \big( j , i  \big)}{v_j \big( t \big) }    \equiv   \frac{\dot{\mathcal{P}} \big( j , i , \dot{M}, u_j , 1 \big)}{v_j \big( t \big) }   =   \frac{\mathcal{P} \big( j , i , \ddot{M}, u_j , 1\big)}{v_j \big( t \big) }   =   u^{\mathrm{*}}_j \big( t \big) \ddot{M \big( t \big) }         \text{ } \text{ , } 
\end{align*}

\noindent in the term appearing before the summation over $i \neq j$. The first time derivative of $M$ has the block representation,

\[
\frac{\partial}{\partial t} M \big( t \big)  \equiv \frac{\partial}{\partial t} M   \equiv  \dot{M}  \big( t \big)   \equiv 
 \begin{bmatrix}
 \frac{\partial}{\partial t} M_{11}  & \frac{\partial}{\partial t} M_{12}
\\ \frac{\partial}{\partial t} M_{21}  & \frac{\partial}{\partial t} M_{22}  \text{ }  
  \end{bmatrix}  \equiv  \begin{bmatrix}
  \dot{M_{11}}  &  \dot{M_{12} }
\\  \dot{ M_{21} } & \dot{M_{22}}  \text{ }  
  \end{bmatrix} \text{ } \text{ , } 
\]

\noindent for $\dot{M_{11}} \equiv \dot{M_{11}} \big( t \big)$, $\dot{M_{12}} \equiv \dot{M_{12}} \big( t \big)$, $\dot{M_{21}} \equiv \dot{M_{12}} \big( t \big)$, and $\dot{M_{22}} \equiv \dot{M_{22}} \big( t \big)$.

\section{References}

\noindent [1] Balian, R. Random matrices and information theory. \textit{Il Nuovo Cimento B} \textbf{57}: 183-193 (1968).

\bigskip

\noindent [2] Lin, Z., Ramezani, H., Eichelkraut, T., Kottos,T., Cao, H., and Christodoulides, D.N. Unidirectional Invisibility induced by PT-Symmetric Periodic Structures. \textit{Phys. Rev. Lett.} \textbf{106} (2011).

\bigskip

\noindent [2] Cotler, J.S. et al. Black holes and random matrices. \textit{Journal of High Energy Physics} \textbf{1705}: 118 (2017).

\bigskip

\noindent [3] Marchenko, V.A., Pastur, L.A. Distribution of eigenvalues for some sets of random matrices. \textit{Mat. Sb. (N.S.)} \textbf{114}(4): 507-536 (1967). 

\bigskip

\noindent [4] Movassagh, R. Eigenvalue Attraction. \textit{Journal of Statistical Physics (Springer US)}, \textbf{162}:3, 615-643 (2016).

\bigskip

\noindent [5] Mostafazadeh, A. Spectral Singularities of Complex Scattering Potentials and Infinite
Reflection and Transmission Coefficients at real Energies. \textit{Phys.Rev.Lett.}. \textbf{102}:220402 (2009).

\bigskip

\noindent [5] Nelson, D.R. Biophysical Dynamics in Disorderly Environments. \textit{Annu. Rev. Biiophys.} \textbf{41}: 371-402 (2012).

\bigskip

\noindent [6] Pastur, L.A. On the spectrum of random matrices. \textit{Theoretical and Mathematical Physics} \textbf{10}(1): 67-74 (1972).

\bigskip

\noindent [7] Tao, T., Vu, V. Random Matrices: The Circular Law. \textit{Communications in Contemporary Mathematics} \textbf{10}(2): 261-307 (2008). 

\bigskip

\noindent [8] Tao, T. Poincaré’s legacies: pages from year two of a mathematical blog \textit{American Mathematical Society} (2009).

\bigskip

\noindent [9] Ticozzi, F., Viola, L. Stabilizing entanbled states with quasi-local quantum dynamical semigroups. \textit{Phil. Trans. of the R. S.} \textbf{370}: 5259-5269 (2012).

\bigskip

\noindent [10] Xeginer, Y. The Expected Norm of Random Matrices. \textit{Combinatorics, Probability, and Computing} \textbf{9}(2): 149-166 (2000).

\end{document}